\documentclass{vgtc}                          

\ifpdf
  \pdfoutput=1\relax                   
  \usepackage{amsmath}
  \usepackage{graphicx}                
  \DeclareGraphicsExtensions{.pdf,.png,.jpg,.jpeg} 
\else
  \usepackage{graphicx}                
  \DeclareGraphicsExtensions{.eps}     
\fi%

\graphicspath{{figures/}{pictures/}{images/}{./}} 

\usepackage{microtype}                 
\PassOptionsToPackage{warn}{textcomp}  
\usepackage{textcomp}                  
\usepackage{mathptmx}                  
\usepackage{times}                     
\usepackage{cite}                      
\usepackage{tabu}                      
\usepackage{booktabs}                  
\usepackage{multirow}
\usepackage[dvipsnames, table,xcdraw, svgnames]{xcolor}

\usepackage{enumitem}

\onlineid{5350}

\vgtccategory{Research}

\vgtcinsertpkg

\title{Spatio-Temporal Mixed and Augmented Reality Experience Description for Interactive Playback}

\author{Dooyoung Kim\thanks{e-mail: dooyoung.kim@kaist.ac.kr}\\ %
         \scriptsize KAIST KI-ITC ARRC %
 \and Woontack Woo\thanks{Corresponding Author. e-mail: wwoo@kaist.ac.kr}\\ %
      \parbox{1.4in}{\scriptsize \centering KAIST UVR Lab. \\ KAIST KI-ITC ARRC}}

\teaser{
  \centering
  \includegraphics[width=\linewidth]{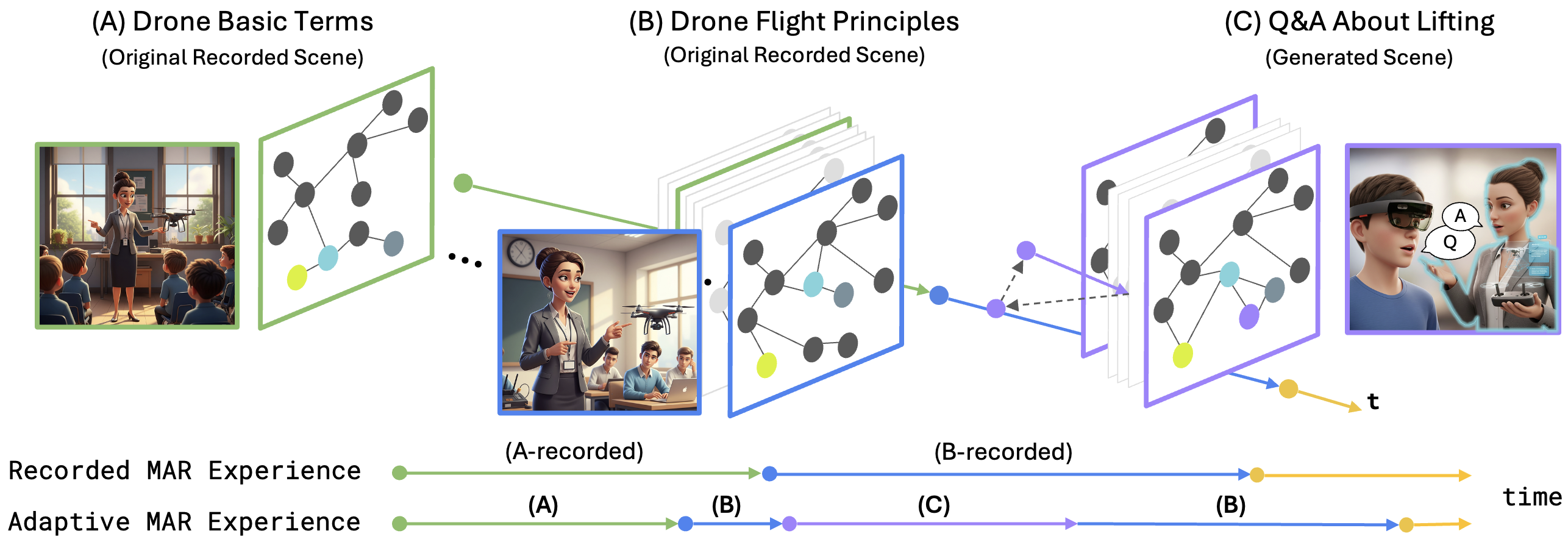}
  \caption{This figure illustrates the concept of a spatio-temporal Mixed and Augmented Reality Experience Description (MAR-ED) for interactive playback. (A, B) An expert's drone tutorial is captured and stored as keyframes within a scene graph, enabling it to be adaptively replayed in different spatio-temporal contexts. The system supports playback speeds tailored to the user's pace and can respond to new interactions, such as (C) user questions or proactive suggestions based on user behavior.}
  \label{fig:teaser}
}

\abstract{
We propose the Spatio-Temporal Mixed and Augmented Reality Experience Description (MAR-ED), a novel framework to standardize the representation of past events for interactive and adaptive playback in a user's present physical space. While current spatial media technologies have primarily focused on capturing or replaying content as static assets, often disconnected from the viewer's environment or offering limited interactivity, the means to describe an experience's underlying semantic and interactive structure remains underexplored. We propose a descriptive framework called MAR-ED based on three core primitives: 1) Event Primitives for semantic scene graph representation, 2) Keyframe Primitives for efficient and meaningful data access, and 3) Playback Primitives for user-driven adaptive interactive playback of recorded MAR experience. The proposed flowchart of the three-stage process of the proposed MAR-ED framework transforms a recorded experience into a unique adaptive MAR experience during playback, where its spatio-temporal structure dynamically conforms to a new environment and its narrative can be altered by live user input. Drawing on this framework, personal digital memories and recorded events can evolve beyond passive 2D/3D videos into immersive, spatially-integrated group experiences, opening new paradigms for training, cultural heritage, and interactive storytelling without requiring complex, per-user adaptive rendering. 
} 


\begin{document}

\firstsection{Introduction}

\maketitle

\begin{figure*}[h]
\centering
\includegraphics[width=\linewidth]{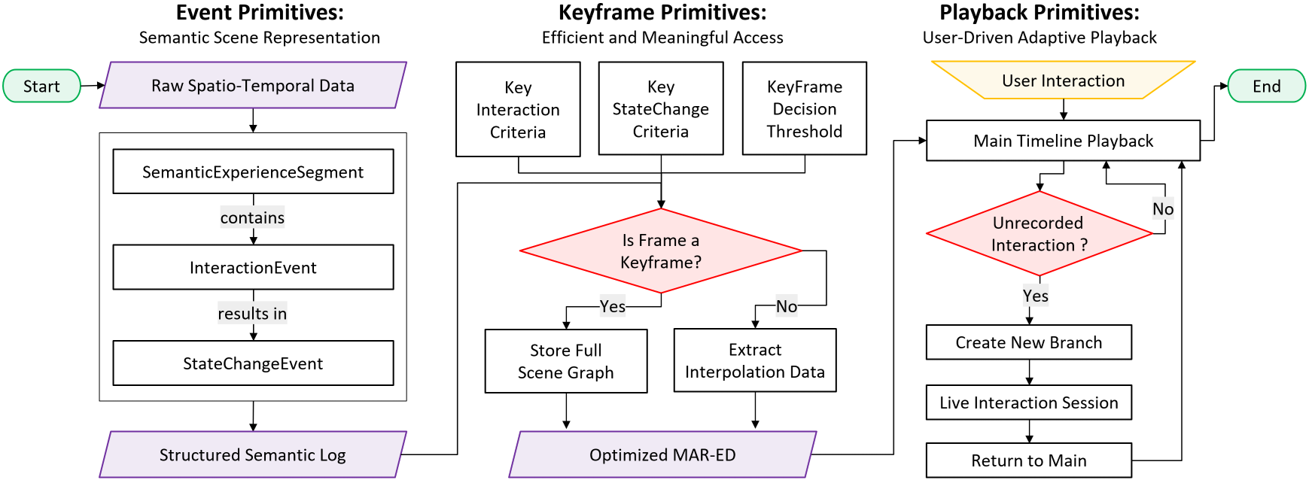}
\caption{The flowchart of the three-stage process of the proposed MAR-ED (Mixed and Augmented Reality Experience Description) framework.}
\label{fig_concept}
\end{figure*}

As immersive technologies continue to blur the boundaries between physical and virtual worlds, the desire to capture and relive memories is evolving beyond the limitations of traditional photos and videos~\cite{kim2025overview}. The future of how we record and re-experience memories points not toward passive viewing, but toward full, interactive immersion. Envision entering an archival recording of a past event, able to converse with the figures within it or interact with key objects to alter outcomes, as imagined in science fiction like Ready Player One~\cite{readyplayerone2018}. This level of engagement requires that recorded memories are no longer static assets to be simply watched, but dynamic worlds to be explored and influenced. However, current spatial media technologies fall critically short of this vision. To enable a future where we can truly step into and interact with our past, a new standard for describing Mixed and Augmented Reality (MAR) experiences is not just beneficial, but essential.

Despite the growing prominence of technologies for capturing our world in 3D, creating a compelling method to replay and interact with these past experiences remains a significant design challenge. The core issue is the lack of a standardized way to describe an experience itself—not just as a collection of visual data, but as a structured narrative of meaningful events, interactions, and spatial relationships. Existing formats are designed to describe assets for presentation~\cite{iso2016mpeg-v, iso2023x3d}, constraining their utility for creating dynamic, interactive, and personalized replays of past events. Despite the growing prominence of technologies for capturing our world in 3D, creating a compelling method to replay and interact with these past experiences remains a significant design challenge. The core issue is the lack of a standardized way to describe an experience itself, not just as a collection of visual data, but as a structured narrative of meaningful events, interactions, and spatial relationships.

To address this gap, our work introduces the Spatio-Temporal Mixed and Augmented Reality Experience Description (MAR-ED), a framework that defines a standard for representing past events for interactive and adaptive playback. This approach transforms a canonical recorded experience into a unique adaptive MAR experience. As illustrated in \autoref{fig:teaser} (Generated using Gemini 2.5 Pro~\cite{GoogleGemini2_5_2025}), this allows a user to begin an adaptive drone tutorial by first engaging with a standard, pre-recorded lesson from an expert on drone principles~(A, B). However, when the user encounters a point of confusion and poses a question, the system dynamically deviates from the recorded path. \autoref{fig:teaser}(C) shows it generates a new, interactive Q\&A session in real-time to address the user's specific query, demonstrating a truly responsive learning environment. Once the question is resolved, \autoref{fig:teaser}'s timeline illustration shows the system seamlessly transitions back to the main tutorial, adapting to the user's pace. The timeline in the \autoref{fig:teaser} visually contrasts this adaptive process against a linear recording, highlighting the non-linear path created by user interaction and the potential for dynamic playback speeds.

This dynamic capability is realized through three foundational sets of descriptive primitives. First, Event Primitives provide the semantic backbone; a \texttt{SemanticExperienceSegment} defines the high-level context, like ``drone assembly,'' while \texttt{InteractionEvent} and \texttt{StateChangeEvent} primitives capture the detailed, meaningful actions and their resulting state changes. Second, Keyframe Primitives identify the most significant moments by evaluating criteria for what constitutes a \texttt{KeyInteraction} or \texttt{KeyStateChange}, using a \texttt{KeyframeDecisionThreshold} to establish anchors for adaptation. Finally, Playback Primitives provide the runtime logic for this interactivity; \texttt{CreateNewBranch} manages the deviation for user queries, \texttt{ReturnToMain} ensures a coherent transition back to the core narrative, and \texttt{TimelineAdaptation} adjusts the experience's timing and spatial layout.

The contributions of this paper are threefold. First, we propose a novel framework, MAR-ED, for describing spatio-temporal experiences based on semantic events rather than raw data. Second, we introduce the concept of event primitives, keyframe primitives, and playback primitives that enable user-driven adaptation and interaction, transforming passive viewing into an active experience. Lastly, we outline a new personal and shared media paradigm that bridges the gap between past recordings and present contexts, enabling more accessible and immersive group experiences.

\section{RELATED WORK}
Existing scene description standards have primarily focused on the efficient, often frame-by-frame, representation of scenes for faithful reproduction. The most relevant standard, MPEG-I Scene Description (ISO/IEC 23090-14)~\cite{iso2022mpeg-i}, excels at the efficient storage and synchronized transmission of timed media assets for presentation. However, this asset-centric model is designed for high-fidelity replay of the scene as it was recorded. It inherently lacks the deep semantic structure required for enabling temporally adaptive playback in a new space and supporting meaningful user interaction with the recorded events themselves.

Furthermore, a significant challenge lies in the adaptation of a recorded experience to a new context. Current standards define animation and timelines rigidly. While powerful for pre-authored content, these mechanisms are not designed to dynamically adapt a recorded spatial trajectory or temporal sequence to a different physical environment. Frameworks like MPEG-21 Digital Item Adaptation~\cite{mpeg21} have addressed content adaptation in a broader sense, and OGC standards like IndoorGML~\cite{ogcIndoorGML} can describe the new spatial context, but a dedicated mechanism to extract semantically significant keyframes from the original experience and use them as anchors for spatio-temporal adaptation is underexplored.

Finally, while the foundations for interactivity are well-established, a higher-level framework for managing interactive narratives is needed. Runtime APIs like OpenXR~\cite{khronos_openxr} provide the necessary interface to hardware for tracking and user input, and standards like MPEG-V~\cite{iso2016mpeg-v} can describe the capabilities of interaction devices. However, these standards provide the low-level pipes for interaction data, not the protocol for handling its consequences within a recorded narrative. A general-purpose state machine language like W3C's SCXML (State Chart XML)~\cite{w3cSCXML5} offers a model for event-driven state transitions, but it is not specialized for the unique challenge of branching a spatio-temporal media experience and then returning to the canonical timeline. A new descriptive capability is required to define how to respond to user interactions and manage the flow between the original recorded path and new, user-driven event branches.

\section{METHODOLOGY}
Mixed and Augmented Reality Experience Description (MAR-ED) is built upon three foundational pillars, each represented by a set of descriptive primitives designed to capture, structure, and replay an experience. \autoref{fig_concept} shows the overall flow of a spatio-temporal experience recording with MAR-ED and adaptive playback of the MAR experience. There are three primitives for the MAR-ED. In the first stage, raw spatio-temporal data is taken as input and transformed into a structured semantic log through SemanticExperienceSegment, InteractionEvent, and StateChangeEvent. The second stage filters the structured log. Based on KeyInteraction and KeyStateChange criteria, along with the KeyframeDecisionThreshold, the significance of each frame is assessed to generate keyframe-based experience data (Keyframed MAR-ED) that contains only the most crucial moments. In the final stage, user-driven adaptive playback is performed based on the distilled data. The experience is replayed along a main timeline, but when a user interaction occurs, a new branch is created (Create New Branch) to respond in real-time. Once the interaction is complete, the system seamlessly transitions back (Return to Main) to the original timeline, providing an uninterrupted experience.


\subsection{Event Primitives: Semantic Scene Representation}
These primitives form the semantic backbone of a recorded experience by structuring it as a narrative of meaningful events rather than a stream of raw data. The relationship between them is hierarchical: a \texttt{SemanticExperienceSegment} acts as a high-level container that provides overall context (the “why"), \texttt{InteractionEvent} primitives detail the specific actions or verbs that occur within that segment (the “what"), and \texttt{StateChangeEvent} primitives document the resulting changes to users and objects, providing descriptive details for those actions (the “how"). Together, they build a rich, interpretable representation of the past event.

\begin{itemize}
    \item \texttt{SemanticExperienceSegment:} This primitive functions as a high-level narrative container, defining a distinct, goal-oriented chapter of the overall experience. It encapsulates the segment's scope, including its start and end times, its primary purpose or task (e.g., “drone lifting method”), and the key participants and objects involved. Grouping related events under a single semantic umbrella provides the necessary context to interpret the purpose of individual actions and state changes that occur within that segment.
    \item \texttt{InteractionEvent:} This primitive serves as the “verb” of the experience, capturing the specifics of a single, meaningful action performed by a user. It provides a structured log that details the connection between an actor and an object, containing not just the type of action (e.g., \texttt{grasp}, \texttt{press}) but also the semantic context of the interaction. This includes the state of the interacted object before and after the action (e.g., its relational change from being \texttt{on(Table)} to \texttt{in(Hand)}), and a precise timestamp. This detailed, event-centric log is fundamental for enabling queries about past actions and for triggering responses during an interactive playback.
    \item \texttt{StateChangeEvent:} This primitive documents the significant changes in the properties of users or objects, often occurring as a direct consequence of an \texttt{InteractionEvent}. It captures the continuous or discrete changes in state, providing the data needed for accurate visualization. This goes beyond simple position to include orientation, posture for users, and intrinsic properties for objects (e.g., a light's \texttt{is\_on} status changing from \texttt{false} to \texttt{true}). For example, during a grasp interaction, this primitive would log the trajectory and transformational changes of both the user's hand and the object from the start of the action to its completion, effectively documenting the outcome of the interaction.
\end{itemize}

\subsection{Keyframe Primitives: Efficient and Meaningful Access}
Not all events hold equal importance. This set of primitives defines a mechanism to distill the continuous flow of events into a discrete set of keyframes, which serve as semantic anchors for efficient storage and adaptive playback. This is achieved by first establishing the criteria for what constitutes a significant \texttt{KeyInteraction} or \texttt{KeyStateChange}. The \texttt{KeyframeDecisionThreshold} then acts as a configurable filter that applies these criteria, allowing the system to record experiences tailored to specific purposes by flagging only the most relevant moments as keyframes.

\begin{itemize}
    \item \texttt{KeyInteraction:} It is identified based on a set of criteria that distinguish semantically significant actions from minor movements within a recorded experience. The primary factor is action semantics, where goal-oriented behaviors such as grasping an object, activating a device, or giving an item to another person are prioritized over ambiguous gestures. The significance of the interacted object is also crucial; an interaction with a primary task-related tool is more significant than one with a background element. Furthermore, an action's importance is elevated if it contributes to narrative progression, such as picking up a key to open a door, which advances the experience to a new stage. Finally, the social context is a key determinant, with interactions involving or directed at another user, like a handshake, being inherently more significant than solitary actions.

    \item \texttt{KeyStateChange}: It is defined by criteria that identify meaningful and abrupt shifts in the state of a user or object, often resulting from a \texttt{KeyInteraction}. These criteria include the magnitude and velocity of physical change, where large or sudden movements, such as a user standing up from a chair or an object being quickly moved, are considered key. More importantly, a change in semantic relationships is a critical indicator; this occurs when an object's logical connection to its environment is altered, for instance, by being placed inside a container, lifted off a surface, or brought into close proximity with another key object. Lastly, a change in an object's intrinsic state, such as its power turning on or off, or a door opening or closing, constitutes a definitive key state change.

    \item \texttt{KeyframeDecisionThreshold:} This is a constant value between \texttt{0} and \texttt{1} that sets the criteria for what qualifies as a keyframe, based on the information from \texttt{KeyInteraction} and \texttt{KeyStateChange} primitives. By adjusting this threshold, it becomes possible to record MAR Experiences tailored to specific objectives. A value of \texttt{0} signifies that every frame is recorded as a keyframe, whereas a value of \texttt{1} signifies that no keyframes are recorded at all.
\end{itemize}

\subsection{Playback Primitives: User-Driven Adaptive Playback}
This component provides the runtime logic that transforms a static recorded experience into a dynamic and interactive adaptive MAR experience. The primitives work in concert to manage the session's flow: \texttt{TimelineAdaptation} provides the foundational adjustment of the experience's timing, adapting the pace to the user's context and needs. Building on this adapted baseline, \texttt{CreateNewBranch} and \texttt{ReturnToMain} collaboratively handle the lifecycle of live user interactions, managing the deviation from and seamless return to the core recorded narrative. These primitives are the engine that drives the interactive nature of the playback.
\begin{itemize}
    \item \texttt{TimelineAdaptation:} This primitive is responsible for dynamically adjusting the temporal structure of the recorded experience to adapt to the user's pace and context. The primary process is temporal scaling, which modifies the duration of recorded events. For example, the playback speed of a tutorial segment can be slowed down after a user asks a question, providing them with more time to apply their new understanding. This ensures the experience feels responsive and is tailored to the individual's learning and experience curve, rather than adhering to a rigid, pre-recorded timeline.
    \item \texttt{CreateNewBranch:} This primitive handles live user interactions that deviate from the original recorded narrative. It captures and interprets input from the playback user (such as gestures, speech, or gaze) and translates it into a recognized new event or intent. Upon recognizing a valid interaction, this primitive generates a new event sequence, creating a narrative branch that diverges from the main timeline. For instance, if a user speaks to a recorded avatar, this primitive initiates a “ask about lifting drone" event, temporarily pausing the original sequence of events to allow for the new interaction.
    \item \texttt{ReturnToMain:} This primitive manages the state and logic for concluding a user-created branch and seamlessly transitioning the experience back to the original timeline. It continuously assesses the status of the interactive session, including the active event segment and the user's level of engagement. Once the conditions for returning are met (e.g., the conversation ends), it identifies the most logical point in the original recorded experience to resume playback, ensuring narrative coherence and a smooth transition from the interactive branch back to the main story.
\end{itemize}

\section{DISCUSSION AND CONCLUSION}
The MAR-ED framework represents a paradigm shift from the passive replay of recorded media to an active, co-creative re-interpretation of past experiences. By prioritizing semantic understanding and adaptive logic, it opens up numerous possibilities while also presenting significant technical challenges. The primary advantage of this approach is the creation of deeply personalized and context-aware media. Applications extend far beyond personal memories. In education and training, a surgeon's procedure could be replayed and interactively explored in a trainee's lab. In cultural heritage, historical events could be reconstructed on-site for tourists to engage with. In entertainment, it enables interactive narratives that adapt to the viewer's own home and actions.

However, the practical realization of MAR-ED is contingent on advances in several key technology areas. The creation of a recorded experience requires robust AI-driven semantic scene understanding to automatically identify events, actions, and states from sensor data in real-time. Generating an adaptive MAR experience in turn demands sophisticated user intent recognition to interpret ambiguous inputs, as well as highly accurate SLAM and spatial mapping to reliably anchor and adapt the experience to a new environment. The MAR-ED framework itself is a high-level proposal. Future work must define a concrete data format (e.g., as an extension to glTF), develop a formal ontology for semantic events, and create a reference implementation to validate its feasibility. A tiered approach, defining profiles from simple playback to fully interactive adaptation, could facilitate gradual industry adoption. These challenges underscore that MAR-ED is not just a new file format, but a roadmap for a new class of intelligent, context-aware media.

In this paper, we introduced the MAR-ED, a framework designed to enable the interactive and adaptive playback of past spatio-temporal experiences. By structuring experiences around semantic primitives for events, keyframes, and playback logic, MAR-ED provides a method to transform static, immutable recorded experiences into dynamic, personalized, and adaptive MAR experiences. This approach bridges the gap between past recordings and the user's present context, allowing recorded narratives to be seamlessly integrated into new physical spaces and altered by live interaction. We detailed the core components of the framework, positioned it against existing standards, and discussed its potential applications and technical challenges. While significant research is needed to fully realize its potential, MAR-ED offers a foundational blueprint for the future of personal media, moving us toward a world where our digital memories are not just watched, but truly re-experienced.

\acknowledgments{%
This work was supported by the Institute of Information \& communications Technology Planning \& Evaluation (IITP) grant funded by the Korea government (MSIT) (No. RS-2024-00397663, Real-time XR Interface Technology Development for Environmental Adaptation).
}


\bibliographystyle{abbrv-doi-hyperref}

\bibliography{template}
\end{document}